\documentclass[structabstract]{aa}  
\usepackage{amsmath}
\usepackage{graphicx}
\usepackage{txfonts}
\usepackage{natbib}
\usepackage{subfig}
\usepackage{booktabs}
\usepackage{upgreek}
\usepackage{setspace}
\usepackage{gensymb}
\usepackage{comment}
\usepackage{multirow}
\usepackage{lipsum}
\usepackage{changepage}
\usepackage{float}
\usepackage{appendix}
\usepackage{tabularx}
\usepackage{epsfig}
\usepackage{booktabs,caption,fixltx2e}
\usepackage[flushleft]{threeparttable}

\begin{document}
%

   \title{Modelling the structure and kinematics of the Firework nebula: The nature of the GK Persei nova shell and its jet-like feature.}


   \author{E. Harvey
          \inst{1},
          M.P. Redman\inst{1}, P. Boumis\inst{2}, S. Akras\inst{3}}

\institute{Centre for Astronomy, School of Physics, National University of Ireland Galway, University Road, Galway, Ireland \\
\email{e.harvey2@nuigalway.ie, matt.redman@nuigalway.ie} 
\and{National Observatory of Athens, IAASARS, I. Metaxa \& V. Pavlou, Penteli, GR-15236 Athens, Greece} 
\and{Observat\'orio do Valongo, Universidade Federal do Rio de Janeiro, Ladeira Pedro Antonio 43 20080-090 Rio de Janeiro, Brazil}
}
    
   \date{\today}



  \abstract
{}
{The shaping mechanisms of old nova remnants are probes for several important and unexplained processes, such as dust formation and the structure of evolved star nebulae. To gain a more complete understanding of the dynamics of the GK Per (1901) remnant, an examination of symmetry of the nova shell is explored, followed by a kinematical analysis of the previously detected jet-like feature in the context of the surrounding fossil planetary nebula.}
{Faint-object high-resolution echelle spectroscopic observations and imaging were undertaken covering the knots which comprise the nova shell and the surrounding nebulosity. New imaging from the Aristarchos telescope in Greece and long-slit spectra from the Manchester Echelle Spectrometer instrument at the San Pedro M\'artir observatory in Mexico were obtained, supplemented with archival observations from several other optical telescopes. Position-velocity arrays are produced of the shell, and also individual knots, and are then used for morpho-kinematic modelling with the \textsc {shape} code. The overall structure of the old knotty nova shell of GK Per and the planetary nebula in which it is embedded is then analysed.}
{Evidence is found for the interaction of knots with each other and with a wind component, most likely the periodic fast wind emanating from the central binary system. We find that a cylindrical shell with a lower velocity polar structure gives the best model fit to the spectroscopy and imaging. We show in this work that the previously seen jet-like feature is of low velocity.}
{The individual knots have irregular tail shapes; we propose here that they emanate from episodic winds from ongoing dwarf nova outbursts by the central system. The nova shell is cylindrical, not spherical, and the symmetry axis relates to the inclination of the central binary system. Furthermore, the cylinder axis is aligned with the long axis of the bipolar planetary nebula in which it is embedded. Thus, the central binary system is responsible for the bipolarity of the planetary nebula and the cylindrical nova shell. The gradual planetary nebula ejecta versus sudden nova ejecta is the reason for the different degrees of bipolarity. We propose that the $\lq$jet' feature is an illuminated lobe of the fossil planetary nebula that surrounds the nova shell.}


\authorrunning{Harvey et al.}
\titlerunning{Kinematics of GK Per's Nova shell and a solution to the jet-like feature}
\maketitle

\section{Introduction}

With a proximity of 470 pc \citep{mclaughlin,Harrison:2013aa}, GK Per (1901) is a nearby, historic, and spectacular post-nova source.
As the nearest and brightest of only two classical nova remnants observed within a planetary nebula to date, the other being 
V458 Vul \citep{Wesson458,Roy:2012aa}, 
it offers the best chance to study the evolution of nova explosion debris within a 
planetary nebula and thus potentially aids the understanding of both types of object. 

A classical nova event is the 
result of thermonuclear runaway on the surface of a white dwarf accreting from, typically, a main sequence or a late G- or K-type 
star \citep{warner}. The accreted shell is ejected, at velocities ranging from 5x$10^2$ to typically $<$ 5x$10^3$ km s$^{-1}$ \citep{BodeNova}, once a critical pressure is reached 
at the core-envelope interface. Before ejection, mixing occurs between the accreted envelope and the white dwarf core through convection, leading to 
a heavy-element enrichment of the roughly solar composition envelope \citep{Casanova:2010aa}.
Dwarf novae, which are also exhibited by the GK Per system, result from an instability in the accretion disk surrounding the white dwarf star; the instability is caused by a disturbance in the 
magnetic field of the white dwarf leading to a brightening of 2-6 magnitudes \citep{osakiDN}. These 
events accelerate winds to the order of 1-6x$10^3$ km s$^{-1}$ \citep{dnvelone,dnvel3,dnvel2}, generally faster than classical nova winds. Dwarf novae are important in terms of 
accretion disk physics as during their rise and fall the majority of the emission is from the disk \citep{osakiDN}.

The onset and progression of thermonuclear runaway in the surface envelope of post-enriched accreted hydrogen (and/or helium) is very 
difficult to explain \citep{Casanova:2010aa}. However, detailed observations of the outflow allow for estimates of the 
total mass and abundances of the heavy elements ejected, as is explored in \cite{helton2011}. 
This provides constraints on the mechanism and efficiency of dredge-up from the underlying white dwarf. 
A full understanding of this process would greatly improve inputs into models for nuclear 
reactions that follow the thermonuclear runaway, which would improve the assessment of ejecta masses and composition \citep{Casanova:2010aa}.

As noted by \cite{ederoclitethesis}, and previously by others (e.g. 
Ringwald et al. 1996 and references therein), that old novae are often neglected after their 
explosive lightcurves have reached quiescence. An important possible consequence 
of these systems is that they are candidates for type Ia supernova progenitors \citep{CNassn1a} along with their recurrent 
counterparts [e.g. T. Crb as discussed in \cite{growingWD}]. Old nova shells give insight into dust production, 
clumping, and ISM dispersal mechanisms, thus they should be understood at all evolutionary phases. 

There are many unknowns surrounding the origin of the morphology of nova shells 
such as whether their ejection is spherically uniform or 
intrinsically bipolar, (e.g. \cite{porterasphericity,lloydshaping}). There is recent evidence in the case of V339 Del that shaping of ejected 
material happens very early on; non-sphericity is evident in the first days after outburst \citep{Schaefer:2014aa}. The common envelope phase is thought to play a major role in the 
shaping of nova remnants, and planetary nebulae alike \citep{balick_shaping,Nord}. Slower nova events (regarding both photometric evolution and ejection velocities) are believed to have stronger deviations from 
spherical symmetry. This hints at the importance of the role the common envelope phase plays, when the time spent by the binary in the envelope is 
longer than their orbital period. Shaping is also expected from interaction with prior ejecta, which would also be non-spherical owing to various similar processes to the nova ejecta. 
There is observational evidence of dependence of 
axial ratio on speed class that can be seen in Fig. 8 of \cite{slavin}. 

As an old, bright, and close nova shell with published kinematics, GK Per is an ideal object to study 
and enrich our knowledge regarding the mechanisms at work in a nova system, see Fig. 1. The 1901 GK Per nova event was possibly its first \citep{Bode04} and 
it has emerged as a well-studied and peculiar object. The central system has changed state, in line with 
hibernation theory \citep{hibernation} although sooner than the theory predicts. 
It was the first classical nova remnant discovered in X-rays \citep{balogel} and non-thermal radio emission \citep{Seaquist89}, implying 
an interaction with pre-existing material surrounding the system. \cite{bode87} first observed that a probable planetary nebula surrounds GK Per and thus is a likely 
candidate for the preexisting material detected in radio and X-rays. The spectacular superluminal light-echos by Wolf, Perrine and Ritchey in the years directly following 
the nova event (explained by \citealp{Couderc} as the forward scattering of light along dust sheets) reveal an abundance of material in the 
vicinity of the system. The first direct image of the nebulosity associated with the classical nova event was taken in 1916 \citep{Barnard}.

GK Per has the longest period classical nova progenitor binary known to date, see Table \ref{orbchar}, with a carbon deficient secondary star \citep{carboncompanion}. 
After undergoing several dwarf nova outbursts, observed since 1963, the object has been reclassified as an intermediate polar, 
which implies the presence of a strong magnetic field (1-10 megagauss, \citealp{watson85}). The dwarf nova outbursts on GK Per were first observed in 1963 and have a recurrence 
timescale of 3 years and a duration of 2 to 3 months. 
However, strong optical outbursts can be traced back to 1948 in the AAVSO data, once the central system had settled down to its quiescent state. 
Two more outbursts followed the 1948 explosion in quick succession, one in 1949 and another in 1950 \citep{sab83_dngk}. 

The central system is seen drifting through the local environment at about 45 km s$^{-1}$, a value derived from proper motion studies \citep{Bode04}. 
X-ray and radio observations reveal evidence of interaction between the SW quadrant of the shell with 
pre-existing material \citep{Anupama:aa,balogel,gkchandra15}. 
The apparent box-like appearance of the nebula has long been observed (e.g. \citealt{Seaquist89}). In the past arguments for flattening of the southern part of the shell through interaction with 
pre-existing material would not explain the flattened northern part of the shell. 
There have, however, been hints of an intrinsic symmetry to the nova shell such as the prolate structure proposed in \cite{Seittermorph}. 
They suggested a broken prolate structure, a missing southern cap, and an extended northern cap spanning the position angles (P.A.) 130-300$^{\circ}$. 
\cite{lawrencefp} conducted three-dimensional Fabry-Perot imaging spectroscopy of GK Per's nova shell, where they presented channel maps 
and a spatial model.

It had been previously believed that the expanding nova remnant was decelerating at a faster rate, based on 
rigorous analysis of proper motions and radial velocities of individual knots \citep{Liimets:2012aa} (hereafter L12) 
came to the conclusion that the system is decelerating at a rate of at least 3.8 times slower than that derived by \cite{Duerbeck}. 
An eventual circularisation of the shell was proposed in L12. From considerations of the data presented in L12 as well as others, such as \cite{Seittermorph,lawrencefp,tweedy} and \cite{Shara:2012aa}, 
the opportunity to explore the structure and kinematics of the GK Per nova shell was undertaken. 
Knots associated with the shell are expanding almost radially away from the central system and have been 
followed over the decades; high-quality spectroscopy is also available. New data were collected to complement those found in the archives.
GK Per exhibits a distinct arc of emission to the NE of the nova shell that is reminiscent of a jet but whose origin 
has not been settled \citep{Bode04,Shara:2012aa}. This study aims to determine the relationship between the expanding knotty debris, the dwarf nova winds, the nature of the jet-like feature and the planetary 
nebula.

  \begin{table}[htbp]
\caption{GK Per, characteristics of the central binary system}
\begin{tabular}[width=0.3\textwidth]{lc c c c cl}
\toprule
Orb T\tablefootmark{1} & WD T & e\tablefootmark{2} & inc($^{\rm \circ}$)\tablefootmark{3} & $M_{\rm 1}$\tablefootmark{4} & $M_{\rm 2}$\tablefootmark{5}\\
\midrule
1.997 d\tablefootmark{a} & 351 s\tablefootmark{b} & 0.4\tablefootmark{c} $\parallel$ 1\tablefootmark{a} &50-73\tablefootmark{d}&0 .9M$_{\rm \odot}$\tablefootmark{a} & 0.25M$_{\rm \odot}$\tablefootmark{a}\\
 \bottomrule
\label{orbchar}
\end{tabular}
\tablefoot{
\tablefootmark{1}{T = Period}
\tablefootmark{2}{e = ellipticity}
\tablefootmark{3}{inc = inclination}
\tablefootmark{4}{$M_{\rm 1}$ = White dwarf mass}
\tablefootmark{5}{$M_{\rm 2}$ = Mass of companion}
\tablefootmark{a}{\citep{crampton_orbper}}
\tablefootmark{b}{\citep{watson85}}
\tablefootmark{c}{\citep{Kraft64}}
\tablefootmark{d}{\citep{Morales-Rueda:2002aa}}
}
\end{table}

This paper follows the structure outlined here: Observations, both archived and new, are presented in Section 2; in Section 3 the analysis 
of the observations is explained; a discussion follows in Section 4; and conclusions in Section 5.




\section{Observations}
\label{Observations}
\subsection{Imaging}

Using the 2.3 m Aristarchos telescope in Greece, new imaging was collected of GK Per on the 27 
August 2014. 
The observations consisted of two narrowband filters focused on H$\alpha$ and [N~{\sc ii}] with exposures of 1800 s in each filter. 
The seeing was of the order of 2 arcseconds. A 1024 x 1024 CCD detector was used where each 24 $\mu$m square is 
$\equiv$ 0.28 arcsec pixel$^{-1}$ after 2x2 binning. The observations are summarised in Table \ref{observations}(a). The imaging data were reduced and world coordinate system matching completed using standard routines in {\sc iraf} \footnote{{\sc iraf} is distributed by the National 
Optical Astronomy Observatories, which are operated by the Association of Universities 
for Research in Astronomy, Inc., under cooperative agreement with the National Science Foundation.}.

Narrowband filters, essential for deducing nebula structure have the disadvantage 
of missing emission of very explosive sources that is Doppler shifted out of the filter bandpass. The Aristarchos narrowband images (H$\alpha$ and [N~{\sc ii}] ), 
which were combined by scaling the flux of field stars to match and then used in the comparison Figures \ref{fig:P-Vssingles}, \ref{fig:P-Vscomplexes}, and \ref{fig:P-Vscomp}, 
cover 
6558.5A-6575.5 $\AA$ and 6579.5-6596.5 $\AA$, respectively, giving a contribution of over 4 $\AA$ from H$\alpha$. The 
information below -200 km s$^{-1}$ is therefore missing, but the information from +560 up to +740 km s$^{-1}$ is accounted for by the 
6548 $\AA$ emission, which is about four times weaker than its 6583 $\AA$ counterpart, but stronger than H$\alpha$. 
For 6583 $\AA$ [N~{\sc ii}] we miss data from -160 to -340 
km s$^{-1}$ and +610. 
From the observations, H$\alpha$ and [N~{\sc ii}] have the strongest emission in the nova shell and together we 
obtain contributions from all knots except those from -200 to -340 km s$^{-1}$. Overall, in the 
-1000 km s$^{-1}$ to +1000 km s$^{-1}$ range we are missing 7\% of the information on the knots comprising the nova shell 
at these constraints. The coverage also shows higher velocity knots down to -1100 km s$^{-1}$ and up to 1500 km s$^{-1}$. 
Owing to the variable emission between the lines covered in the Aristarchos observations and to differences in sensitivity, 
caution is to be exercised 
when judging differences in knots between these observations and those taken with the NOT + Mayall telescopes, see Figures 
\ref{fig:P-Vssingles}, \ref{fig:P-Vscomplexes}, and \ref{fig:P-Vscomp}. The degraded seeing conditions of the Aristarchos telescope observations 
would be expected to mostly compensate for the missing information.

\begin{table*}
\caption{Observations: (a) are newly acquired for this 
work whereas (b) were obtained from data archives.}
 \centering
\begin{tabular}[width=0.5\linewidth]{lc c c c cl}
  		(a)& & & \\
     \toprule
        Date       & Type       & Telescope & Instrument & 	Filter CWL/FW($\AA$)  & P.A.\tablefootmark{*} 	& 	Exp. (sec)\\
  \midrule
        2014-Aug-27        & Imaging       & Aristarchos\tablefootmark{1} - 2.3m	& LN CCD & H$\alpha$ 6567/17& &	1800\\
                & Imaging       & Aristarchos - 2.3m	&LN CCD& [N~{\sc ii}]  6588/17& &	1800\\
        \hline
       2014-Nov-28		& Spectroscopy		& 	SPM\tablefootmark{2}  - 2.12m & MES &	H$\alpha$+[N~{\sc ii}] 6541/90& 45$^{\circ}$ & 1800\\
        		& Spectroscopy		& 	SPM - 2.12m & MES & 	H$\alpha$+[N~{\sc ii}] 6541/90& 30$^{\circ}$ & 1800\\
        		& Spectroscopy		& 	SPM - 2.12m &	MES & [O~{\sc iii}] 4984/60&  45$^{\circ}$ & 1800\\
        		& Spectroscopy		& 	SPM - 2.12m &	MES & [O~{\sc iii}] 4984/60& 30$^{\circ}$ & 1800\\
		\hline
		2015-Mar-27		& Spectroscopy		& 	SPM - 2.12m & MES &	H$\alpha$+[N~{\sc ii}] 6541/90& 9$^{\circ}$ & 1800\\
		\hline
		2015-Mar-28		& Spectroscopy		& 	SPM - 2.12m &	MES & H$\alpha$+[N~{\sc ii}] 6541/90& 9$^{\circ}$  & 1800\\
				& Spectroscopy		& 	SPM - 2.12m &MES & 	[O~{\sc iii}] 4984/60& 9$^{\circ}$ & 1800\\
				 \bottomrule
\\
		(b)& & & \\
    	\toprule
        Date       & Type       & Telescope	&  Instrument &	Filter/grism CWL/FW($\AA$) & P.A. 	& 	Exp. (sec)\\
      \midrule
      	1995-Nov-8        & Imaging       & HST\tablefootmark{3} - 2.4m	& WFPC2 & F658N 6591/29 & &	1400\\
        1997-Aug-1        & Imaging       & HST\tablefootmark{3} - 2.4m	& WFPC2 & F658N 6591/29 & &	1200\\
        \hline
       2007-Sep-3/5        & Imaging       & NOT\tablefootmark{4} - 2.5m	& ALFOSC & H$\alpha$ 6577/180& &	3600\\
       	\hline
        2010-Feb-6/7       & Imaging       & KPNO\tablefootmark{5} - 4m	&CCD Mosaic& H$\alpha$ 6563/80& &	7440\\
        \hline
        2010-Feb-12		& Imaging	  & WISE\tablefootmark{6} - 40cm	&Survey Camera& B3 (12$\mu$m)\tablefootmark{7} & & 4422\\
        \hline
        2007-Sep-3		& Spectroscopy		&	 NOT - 2.5m & ALFOSC	&  g17 6600/500 & 31$^{\circ}$ & 3600\\
        \hline
        2007-Sep-4		& Spectroscopy		&	NOT - 2.5m & ALFOSC	  & g17 6600/500 & 86$^{\circ}$ & 1800\\
        		& Spectroscopy		&	NOT - 2.5m & ALFOSC	 & g17 6600/500 & 107$^{\circ}$ & 1500\\
        		& Spectroscopy		&	NOT - 2.5m & ALFOSC & g17 6600/500 & 312$^{\circ}$ & 1800\\
        \hline
        2007-Sep-5		& Spectroscopy		&	NOT - 2.5m & ALFOSC	 & g17 6600/500 & 49$^{\circ}$ & 1800\\
        		& Spectroscopy		&	NOT - 2.5m & ALFOSC	 & g17 6600/500 & 173$^{\circ}$ & 1800\\
        \hline
        2010-Feb-10        & Spectroscopy       & KPNO - 4m & RC-Spec	& BL400 5500/7000 & 89.7$^{\circ}$ &	9600\\
         \bottomrule
        \label{observations}
    \end{tabular}
 	\tablefoot{
\tablefootmark{1}{Aristarchos Telescope, Helmos Observatory}
\tablefootmark{2}{San Pedro M\'artir Observatory}
\tablefootmark{3}{Hubble Space Telescope}
\tablefootmark{4}{Nordic Optical Telescope}
\tablefootmark{5}{Mayall Telescope, Kitt Peak National Observatory}
\tablefootmark{6}{Wide-field Infrared Survey Explorer}
\tablefootmark{7}{WISE band 3 has a resolution of 6.5$^{\prime\prime}$.}
\tablefootmark{*}{P.A. = Position Angle.}
}
\end{table*}


\subsection{Spectroscopy}
\begin{figure}
\includegraphics[width=9cm]{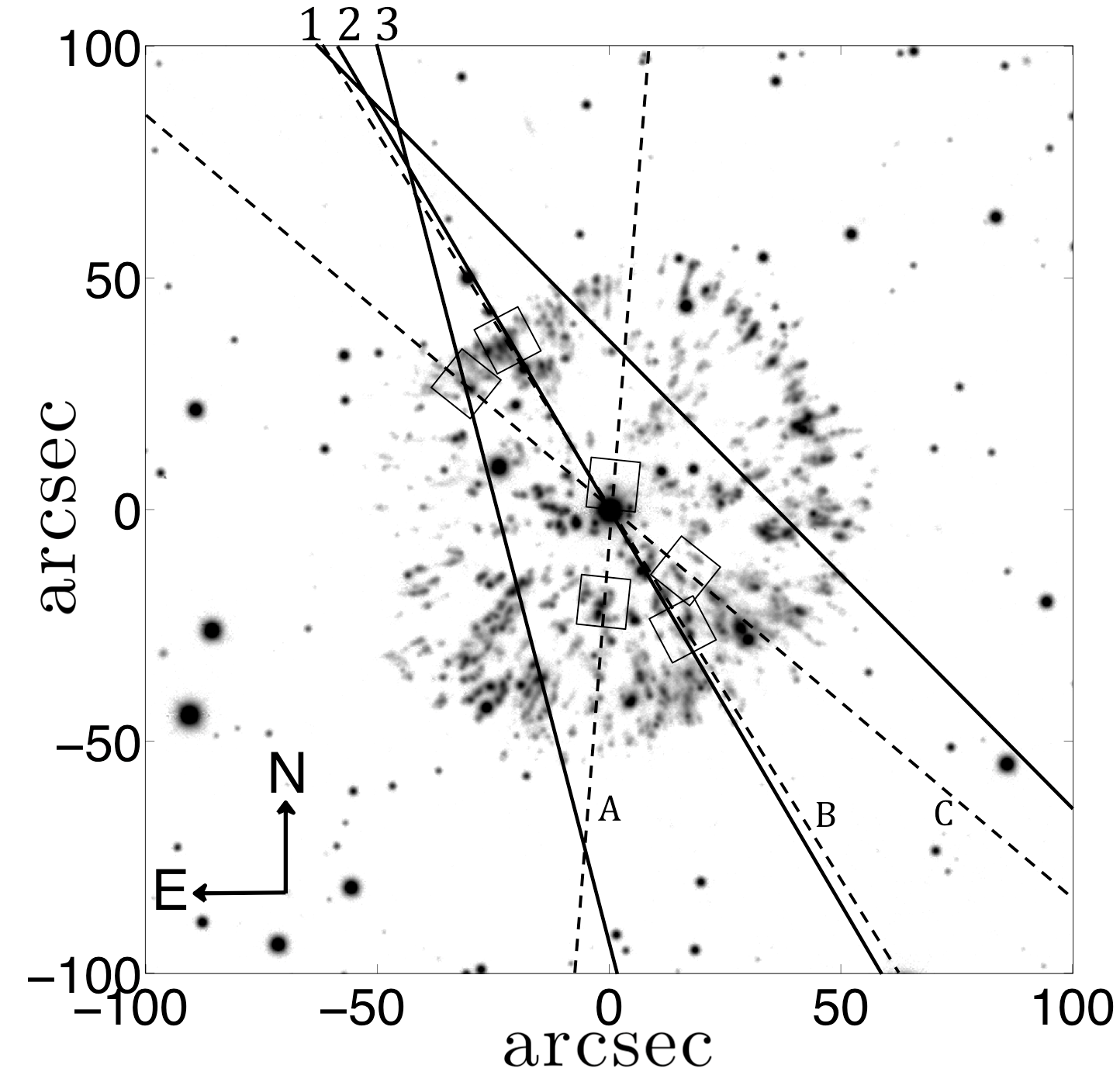}
\caption{Positions of the slits from MES-SPM overlaid on the stacked H$\alpha$ Mayall image first presented in \cite{Shara:2012aa}. The slit marked 1 was placed at a P.A. of 45$^{\circ}$, slit 2 at 30$^{\circ}$ and slit 3 at 9$^{\circ}$. All three slit positions cover the jet-like feature to the NE of the nova remnant as well as sections of the nova shell. Slit 1 \& 2 observations were obtained on 2014-Nov-28 and slit 3 on 2015-Mar-27/28, 
see Table \ref{observations}(a). Slits marked A, B, and C are those at P.A. = 173$^{\circ}$, 31$^{\circ}$, and 49$^{\circ}$, respectively. It 
is from these slit positions that the P-V arrays were simulated in Figures \ref{fig:P-Vssingles}, \ref{fig:P-Vscomplexes}, and \ref{fig:P-Vscomp}. The overlaid boxes cover the knots used in the same figures.}
\label{fig:slits}
\end{figure}

Echelle spectroscopic data were also obtained in order to 
measure the extinction velocities of several interesting knots, to build a more complete view of the remnant, and to examine the jet-like feature. 
These were obtained using the Manchester Echelle Spectrometer (MES) instrument at the San Pedro M\'artir (SPM) observatory in Mexico \citep{MeaburnMES}. The data were collected over the course of two observing periods, 
November 2014 and March 2015, and in three different positions. The slit positions, overlaid on an image of the system in Fig. \ref{fig:slits}, were observed with the instrumentation in its f/7.5 configuration. A Marconi 2048x2048 CCD was used with a resultant spatial resolution $\equiv$ 0.35 arcsec pixel$^{-1}$ after 2x2 binning was applied during the observations with the {\raise.17ex\hbox{$\scriptstyle\sim$}}6' slit. Bandwidth filters of 90 and 60 $\AA$ were used to isolate the 87$^{th}$ and 113$^{th}$ orders containing the H$\alpha$+[N~{\sc ii}] $\lambda\lambda$6548, 6583 and  [O~{\sc iii}] $\lambda$5007 nebular emission lines. 


\subsection{Archival data}
Archival imaging and spectroscopy from different epochs and telescopes were obtained from the NOAO, NOT, IRSA, and  MAST databases. 
The archival observations are from the Hubble Space Telescope, the WISE satellite, the Nordic Optical Telescope and the Mayall telescope, 
and are summarised in Table \ref{observations}(b). The principal data sets used were Mayall and HST observations reported by  \cite{Shara:2012aa} (hereafter S12), NOT data by L12, and 
WISE  nova data first studied in \cite{evanswise}. Further details and analyses of these data are contained in these references. Here the data were collated and reduced in the same manner as the more recent MES and Aristarchos observations in {\sc iraf}.

Archival echelle spectroscopy bi-dimensional line profiles or position-velocity (P-V) arrays were generated for the six ALFOSC position angles listed in Table \ref{observations}(b). For each P.A. there were between four and eight 
individual knots or knot complexes, for which P-V information could be subtracted. 
The P-V arrays were created by comparing rest wavelengths to the position of the strongest spectral lines of wavelength calibrated spectra. In this way velocity scales were 
made available for subsequent analysis using the Doppler-redshift relation.

%
\section{Analysis}
\label{Analysis} 

The P-V arrays were mostly created 
from archival ALFOSC data and for one position angle they are complemented by more recent echelle spectra from MES-SPM (P.A. = 30$^{\circ}$ and 31$^{\circ}$, respectively, see Table \ref{observations} and Fig. \ref{fig:slits}),
allowing for constraints to be made on spatio-kinematics of individual systems. The P-V arrays were simulated using the 
morpho-kinematic code {\sc shape} \citep{shapenew}
\footnote{A full discussion on {\sc shape} modifiers can be found at $\mathrm{http://bufadora.astrosen.unam.mx/shape/}$} 

Modelling with {\sc shape} is carried out on different spatial scales. The nova shell as directly modelled from observations of the knots 
is described first in Sect. 3.1. Section 3.2 explores the structure of the overall distribution of the knots and Sect. 3.3 reveals new results on the 
jet-like feature associated with the remnant.  

\subsection{Modelling individual knots}

\begin{figure*}
\centering
\includegraphics[width=17cm]{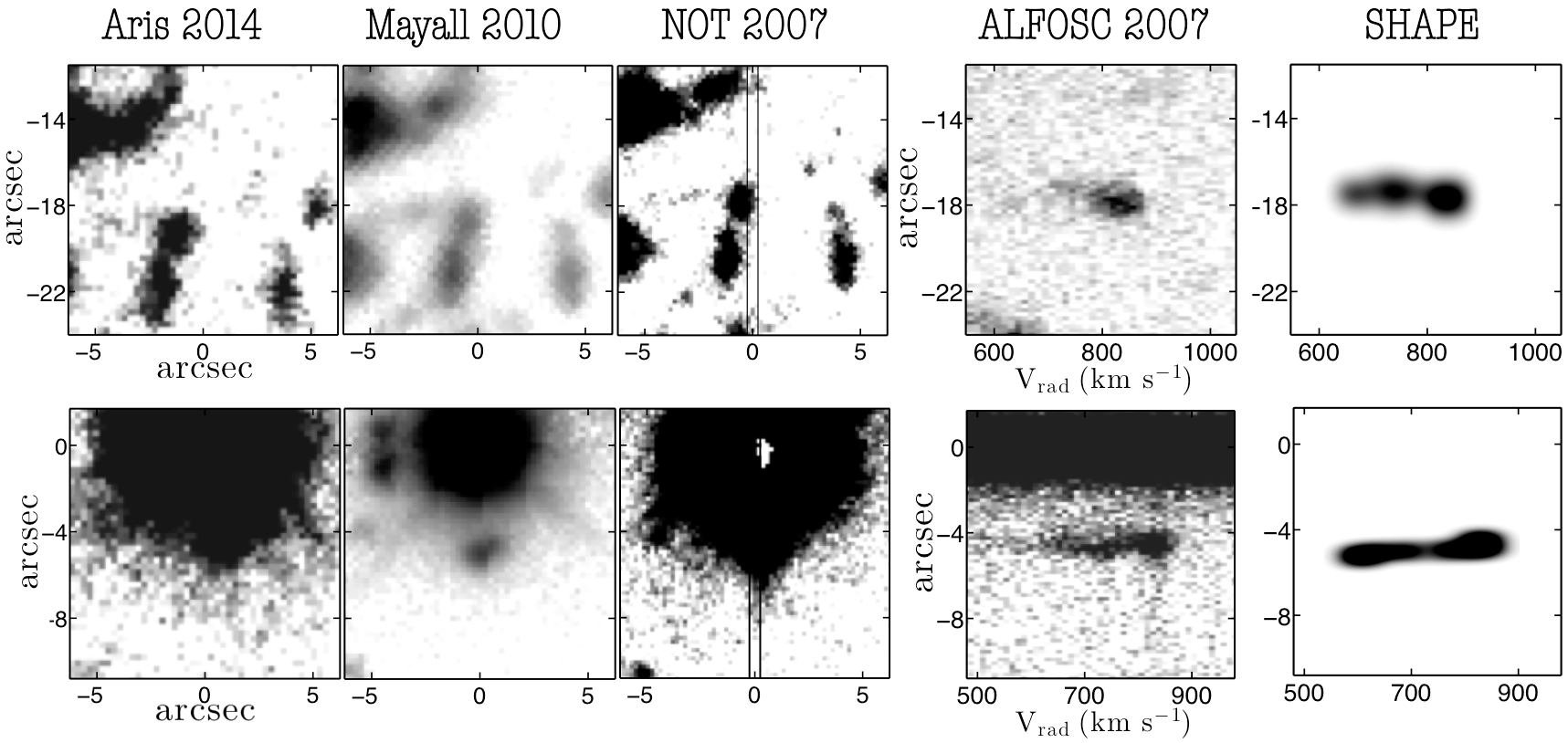}
\caption{Examples of single knots progressing over time (2007-2014) and their corresponding P-V array and morpho-kinematic {\sc shape}. Spatial extent is 12.5 arcseconds on each side, and the velocity ranges are 500 km s$^{-1}$. On the NOT panels the overlaid slit width and position is shown. The Aristarchos and NOT panels in the bottom row are contaminated by the central star leading to a very strong contrast between knots and star. Here, and in Figures \ref{fig:P-Vscomplexes} and \ref{fig:P-Vscomp} the displayed knots are positioned along the observed slit axis highlighted in Fig. \ref{fig:slits}, where the top row corresponds to a slit position angle of 49$^{\circ}$ and the bottom row to 173$^{\circ}$.  The spatial axis distance from the central star is positive towards the top of the slit and negative towards the bottom.}
\label{fig:P-Vssingles}
\end{figure*}

\begin{figure*}
\centering
\includegraphics[width=17cm]{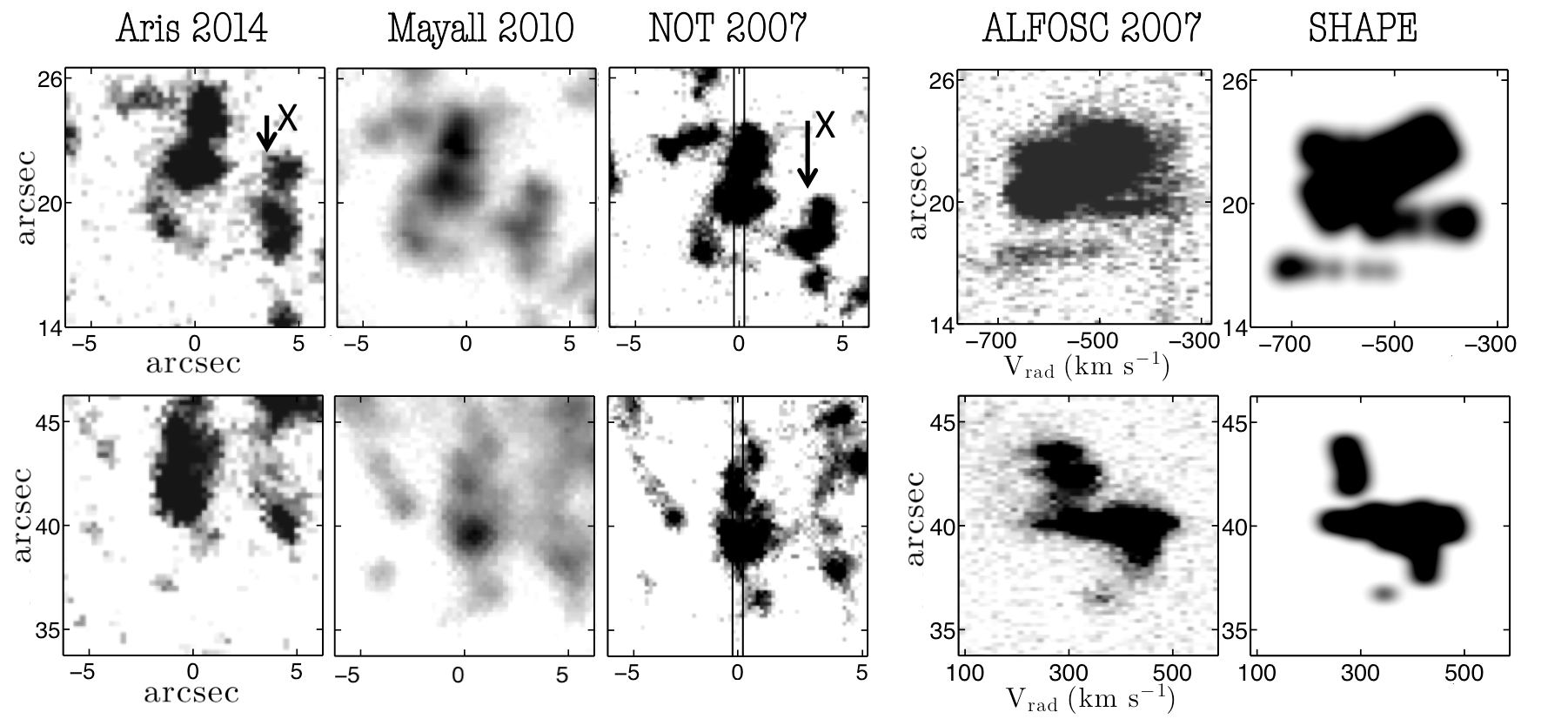}
\caption{Same as in Fig. \ref{fig:P-Vssingles} except with examples of more than one knot in each image. Spatial extent is 12.5 arcseconds and velocity scales are 500 km s$^{-1}$. 
The top row shows the separation of knots just to the bottom right of the multiple knots covered with the 173$^{\circ}$ slit, the bottom row corresponds to a position angle 49$^{\circ}$. The knot marked $\lq$X' can be seen to break away from the knot covered by the slit next to it in HST images from 1995 and 1997; the corresponding P-V array of this knot shows ongoing separation of knots in this clump.}
\label{fig:P-Vscomplexes}
\end{figure*}

\begin{figure*}
\centering
\includegraphics[width=18cm]{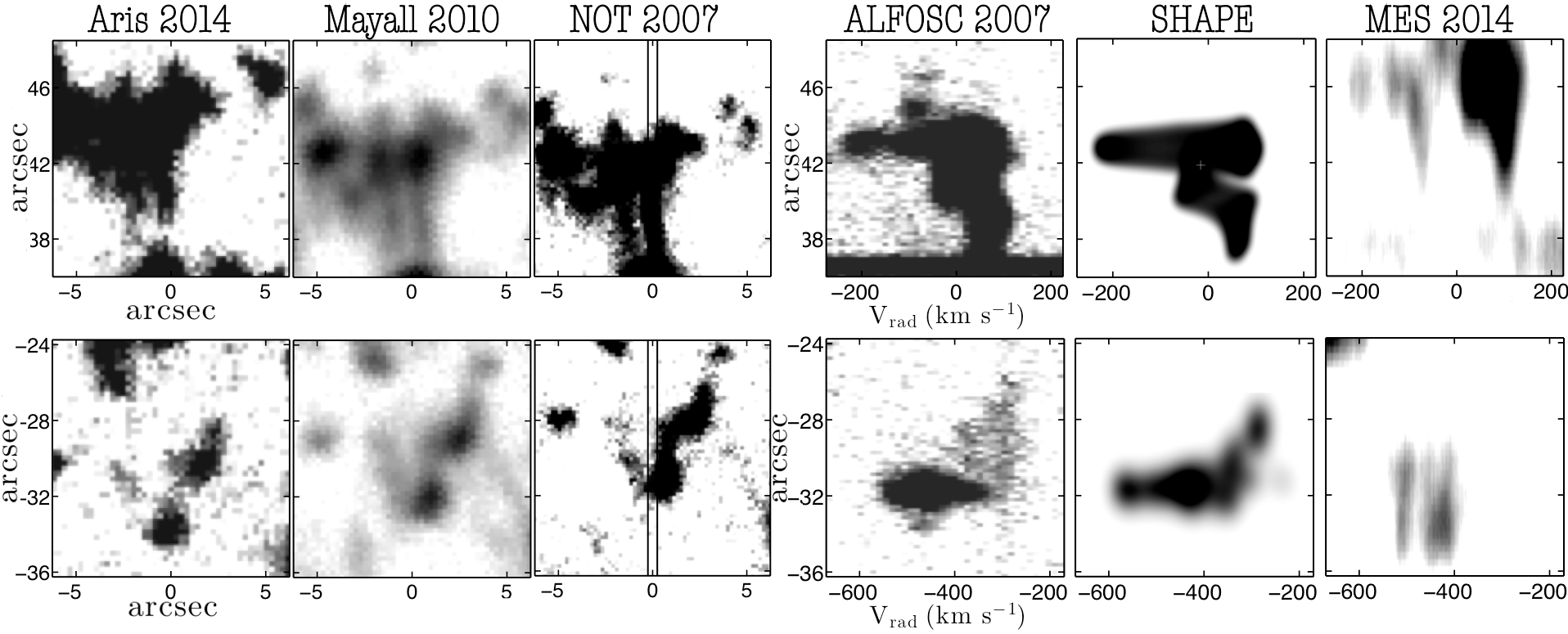}
\caption{Same as in Figures \ref{fig:P-Vssingles} and \ref{fig:P-Vscomplexes} except with additional epoch P-V information. The knots here are along an axis of symmetry of the nova shell (slit position angles of 30$^{\circ}$  and 31$^{\circ}$ for the ALFOSC and MES observations respectively). The ALFOSC P-Vs correspond to the 2007 NOT image and the MES , [N~{\sc ii}] 6583.39$\AA$, P-Vs correspond to the Aristarchos observations epoch.}
\label{fig:P-Vscomp}
\end{figure*}

In the first instance, a direct model from 
published data associated with L12 was made. For this initial model, 115 knots (from the online table of L12) with 
the most substantial data were modelled as individual cylinders with expansion velocities, 
distance from the central star, P.A., and inclinations derived using the prescriptions of L12. In order to achieve this, 
each individual knot was modelled as a cylinder with x, y, z positions in arcseconds specified with the translation modifier; the P.A. and inclination 
were replicated using a rotation modifier; and the expansion velocities specified explicitly with the velocity modifier. We note the modelled microstructure in panel (c) Fig. 6. Later a radial velocity component was added to knots to match the ALFOSC P-V arrays in Figures \ref{fig:P-Vssingles}, \ref{fig:P-Vscomplexes}, and \ref{fig:P-Vscomp}.   
Irregularities in knot structure were then simulated by modifying the cylinder primitives to match their respective P-V arrays using bump, stretch, and squish modifiers. For another recent application and further explanation of these {\sc shape} modifiers see \cite{clyner}.

\begin{figure*}
\centering
\includegraphics[width=18cm]{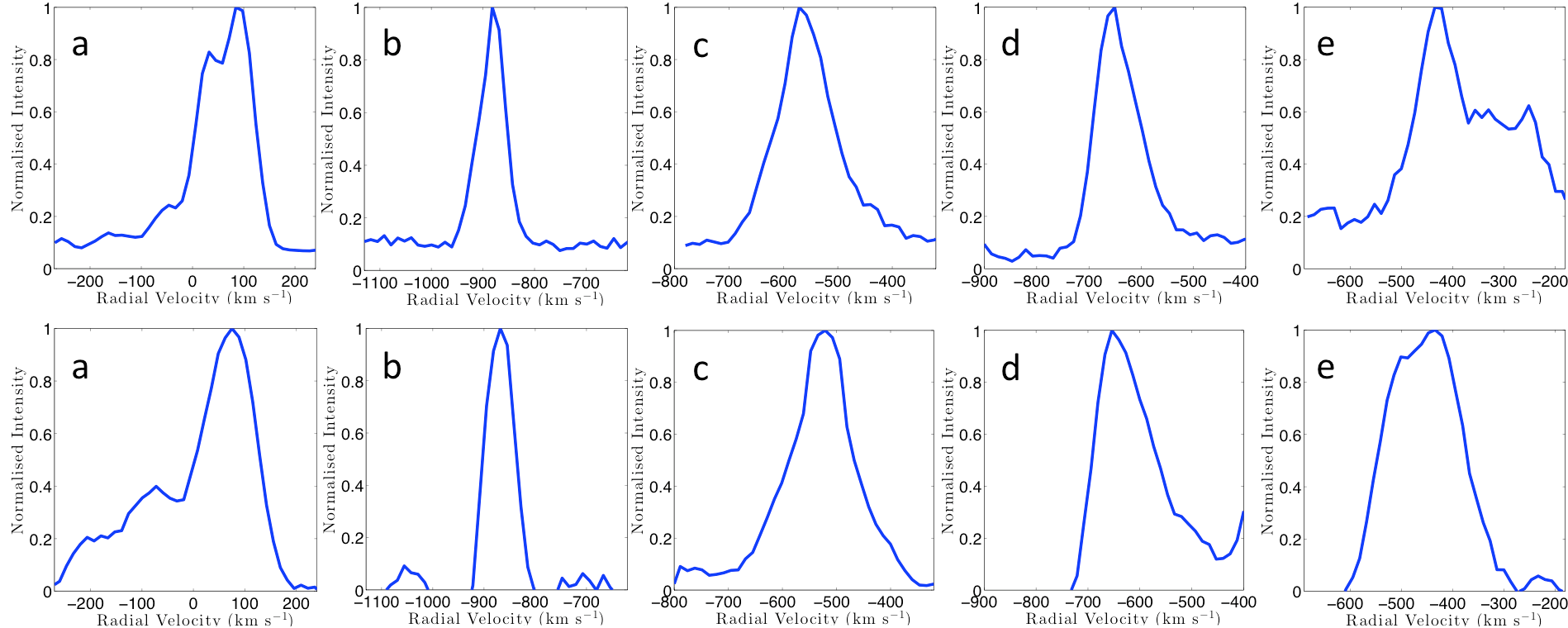}
\caption{Radial profiles of five bright knots from the ALFOSC slit with PA = 31$^{\circ}$ in the top row, and the corresponding MES knots with slit at PA = 30$^{\circ}$, in the bottom row. The knots displayed are seen in their [N~{\sc ii}] 6583.39$\AA$ emission line, knot $\lq$a' corresponds to the top row of Fig. \ref{fig:P-Vscomp} and knot $\lq$e' to the bottom knot of the same figure.}
\label{fig:RV_profs}
\end{figure*}

High-resolution images from several time frames were analysed in order to 
gain further knowledge on the interaction and fragmentation of clumps. 
Variations in the knot morphology and velocity distribution between different clump systems 
can be seen in Figures \ref{fig:P-Vssingles}, \ref{fig:P-Vscomplexes}, \ref{fig:P-Vscomp}, and \ref{fig:RV_profs}. 
The long-slit spectroscopy discussed in L12 was 
re-analysed with a different intent, with the addition of our new data. 
As L12 derived the kinematics based on 
single Gaussian peaks, we aimed to delve further and to examine the extended profiles of individual knots. 
To gain an understanding of the knot kinematics and morphology, spatially oriented P-V arrays were 
created using {\sc iraf} and {\sc Octave} \citep{octave}. The P-V arrays were difficult to reproduce with {\sc shape} owing to the amorphous nature intrinsic to the knots and the complex velocity field in which they 
reside. 

Possible mechanisms were explored to explain the irregularities observed in the individual knot shapes, including 
dynamical instabilities with a 
main contributor thought to be the fast periodic winds produced by the frequent dwarf 
nova episodes. Interestingly, clumping of the ejecta began 
before the dwarf nova episodes had been observed. The initial clumping was most likely due to 
early interacting winds \citep{balogel}.
In Fig. \ref{fig:P-Vssingles} extended velocity-tails of individual knots are apparent; 
these tails appear wavy, which indicates their complex velocity structure. The longest tails 
are closer to the central system in the plane of the sky ( the lower row in Fig. \ref{fig:P-Vssingles} is closer to the central 
star in the plane of the sky than the upper row and thus has a longer tail in velocity space since more velocity information 
is contained in these knots owing to their inclination towards the observer), hinting that their true 
structure is similar 
to those with $\lq$wavy' tails attributed in this work to shaping by dwarf nova winds (see Sect. \ref{discussion}). The longer tails present 
towards the central system was shown previously by L12, see their Fig. 8, where they mention that this could be a projection effect.
Looking at Fig. \ref{fig:P-Vscomplexes}, the first row demonstrates overlapping of knots that can be separated by 
morpho-kinematic modelling, whereas the second row demonstrates clear evidence of a bow-shock and low expansion-velocities also attributable to interaction between knots. 

The difference in the P-V arrays in Fig. \ref{fig:P-Vscomp} over the seven year interval appear significant. The MES data has higher velocity resolution (22 km s$^{-1}$ versus 32 km s$^{-1}$), but lower sensitivity, see Table \ref{observations}. As can be seen in Fig. \ref{fig:P-Vscomp}, top row of panels, 
the ALFOSC P-V array and MES P-V array show differences in position as the proper motion and velocity-structure differences are due to a combination of different sensitivity and resolution, and are also due to a slight (1$^{\rm \circ}$) difference in the slit positions. However, the 1$^{\circ}$ difference in position angle gives a separation of the slits of around 1" at a distance of 60" from the central system, which is within the seeing of both sets of observations. In addition the MES slit is 3.8" wide on the plane of the sky. The pattern of a bow shock around one or more knots remains evident with comparison to the {\sc shape} model, see the bottom row of Fig. 3 and the top row of Fig. 4. Also, in Fig. \ref{fig:P-Vscomplexes}, the knots marked $\lq$X' are seen to undergo shaping changes and bifurcation over the time interval of the presented observations. Tracing these knots back to the HST imaging campaign of S12 they are seen to separate from the knot complex covered by the slit in the same panels as knot $\lq$X'. In Fig. \ref{fig:P-Vssingles} the knot at the upper panel appears to have a tail that is bending back; this knot has a radial velocity of about -650 km s$^{-1}$ and the knot possibly associated with it to the left has a radial velocity of -570 km s$^{-1}$ 31$^{\rm \circ}$ in the ALFOSC long-slit observation and -530 km s$^{-1}$ seven years later in the MES data, see knot $\lq$c' in Fig. \ref{fig:RV_profs} and Table \ref{comparison}, which may be indicative of knot changes due to the complex dynamics of the environment. We note that Fig. \ref{fig:RV_profs} is complemented by Table \ref{comparison} where the velocity widths and signal-to-noise ratios of the observations are shown.

We also note that the analysis of observation leads to a small number of knots (8 out of 115) that have large uncertainties in their deprojected true positions within the population of knots; these uncertainties are caused by large relative errors in one or more components of the deprojected distance estimates, or else are ballistic knots like those discussed in S12. 

Examination 
of the radial velocity distribution with respect to x, y positions of the knots is suggestive of an overall symmetry in the total distribution. The low-resolution channel maps of \cite{lawrencefp} show similar symmetry on reinspection, see Fig. \ref{fig:channelmaps}.

\subsection{Overall distribution} 
The second modelling approach involved testing of basic shapes to fit the overall distribution of the knots 
with x-y positions and radial velocity measurements; the sample size was increased to 148 through the inclusion of knots covered in our new observations, see the table in the Appendix. 
As an example of the versatility of this approach consider a sphere
with a thin shell and the knots added by means of a texture modifier given the 
appropriate filling factor. The positions of the knots can be matched by adjusting their 
position on the shell so that the model and actual P-V arrays agree. 
The sphere can be deformed appropriately to simulate the somewhat box-like appearance of the 
remnant, or it can also be manipulated to have a bipolar, prolate or oblate shape. Also, the size of the structure can be easily modified to see if the same shape can match that of previous imaging campaigns, such as that of \cite{Seaquist89}. Comparison of the brightness distribution between epochs give clues on the sequence of which components and sides started becoming shock illuminated.  However, a sphere primitive was found to be unable to reproduce the channel maps of \cite{lawrencefp}.

\begin{figure*}
\centering
\includegraphics[width=19cm]{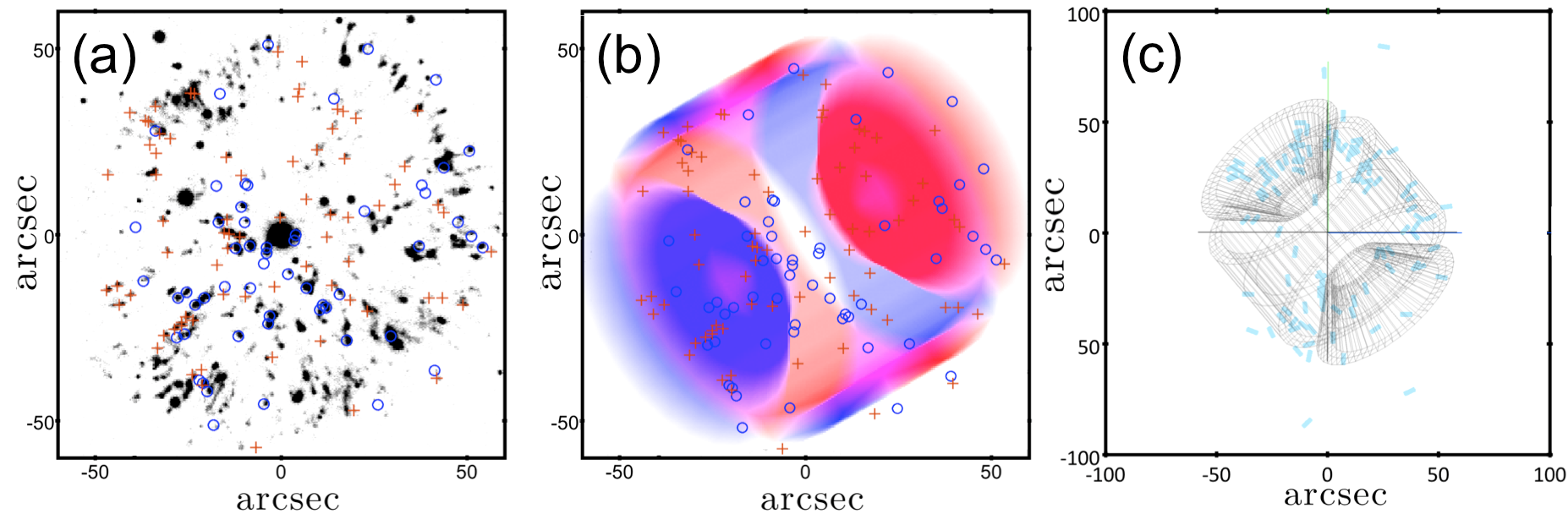}
\caption{Panel (a) shows the red-blue Doppler distribution from observed radial velocities of 148 knots overlaid on an image from the Mayall telescope; north is up and east is to the left. 
Panel (b) shows the fit of the cylinder on the observed radial velocity distribution. 
Panel (c) shows the model fit to the deprojected observations of L12 in the x-z plane. We note the extended ballistic knots and the larger plotted area here.  }
\label{fig:explines}
\end{figure*}

This second approach was then applied to several common nova shell morphologies 
that can be manipulated in the same ways as the sphere. To achieve the best fit an 
inclination, P.A., and expansion velocity were specified (see Table \ref{cylinderfeatures}) implemented using the rotation and 
velocity modifiers in {\sc shape}. To simulate the knots a texture modifier 
was added and the knots were projected radially through the thickness of the cylindrical shell.
Structure was added subsequently to the cylindrical model to account for spatial and Doppler shift 
irregularities. This structure consists of polar cones, see Fig. \ref{fig:explines} and Table \ref{cylinderfeatures},  
the polar features have low equatorial velocities and higher polar velocities. 

\begin{table}
  \begin{threeparttable}
    \caption{Best fit {\sc shape} orientation and velocity model parameters of the GK Per nova shell.}
     \begin{tabular}{llccccccc}
        \toprule
        & P.A. & inc & R$\rm_{in}$ & R$\rm_{out}$ & L & D & V$\rm_{exp}$\\ 
        & ($^{\circ}$) & ($^{\circ}$) & (pc) & (pc) & (pc) & (pc) &  (km s$^{-1}$)\\

        \midrule
                 &     \multicolumn{7}{c}{}     \\
O & 120 & 54 & 0.093 & 0.109 & 0.125 & 0.093 & 1000\\
E & 120 & 54 & 0.027 & 0.091 & 0.027 & 0.080 & 750\\
W & 120 & 126 & 0.027 & 0.091 & 0.027 & 0.080 & 750\\ 

        \bottomrule
             \label{cylinderfeatures}
     \end{tabular}
    \begin{tablenotes}
      \small
      \item The size of the shell features are normalised to the shell's size during the 2007 observations of L12. P.A. = Position Angle, inc = inclination, R$\rm_{in}$ = inner radius of feature, R$\rm_{out}$ = outer radius of feature, L = length, D = distance to central binary, V$\rm_{exp}$ = expansion velocity. With regards to the features O = outer cylinder, E = eastern cone, W = western cone.
    \end{tablenotes}
  \end{threeparttable}
\end{table}

\begin{table}[]
\centering
\caption{Comparison of RV profiles of the five brightest knots in common of the 30 and 31 slits. ALF = ALFOSC on the NOT telescope, MES at San Pedro Martir.}
\label{comparison}
\begin{tabular}{lllllll}
        \toprule
Knot & FWHM & FWHM  & FW10$\%$  & FW10$\%$  & SNR & SNR \\
 & km s$^{-1}$ & km s$^{-1}$ & MES & ALF & MES & ALF \\
 & MES & ALF & & & & \\
         \midrule
a & 112 & 114 & 297 & 214 & 102 & 106\\ 
b & 54 & 59 & 97 & 112 & 47 & 27 \\ 
c & 75 & 120 & 177 & 243 & 35 & 46 \\ 
d & 116 & 92 & 266 & 154 & 26 & 42 \\ 
e & 164 & 106 & 231 & 305 & 81 & 94 \\ 
        \bottomrule
\end{tabular}
\end{table}

\begin{figure*}
\centering
\includegraphics[width=18cm]{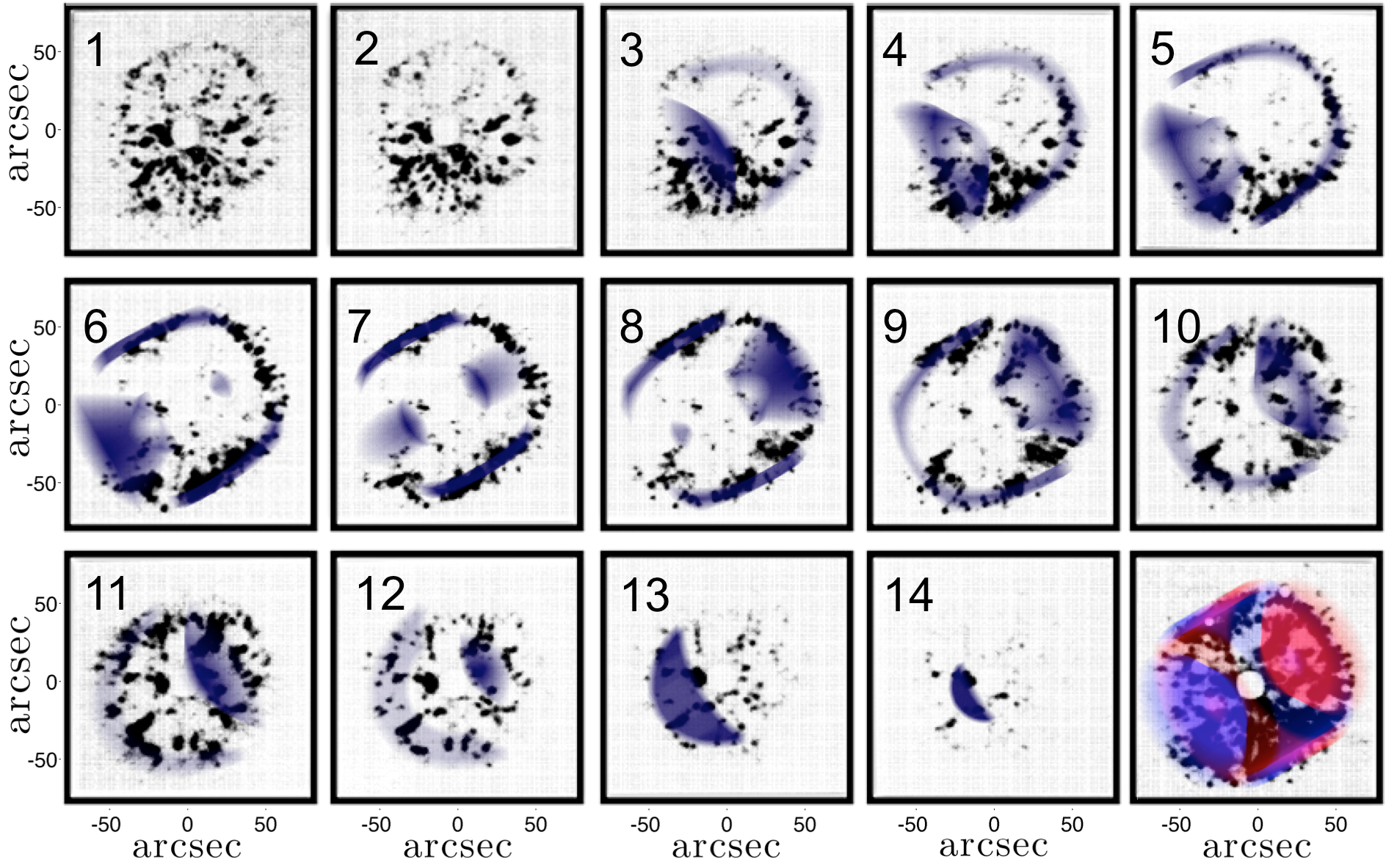}
\caption{Channel maps from \cite{lawrencefp} (dots) compared to our scaled morpho-kinematic 
model (overlaid shadow). Frames 1-14 start at -840 km s$^{-1}$ and go up to 980 km s$^{-1}$, 
the velocities being relative to the [N~{\sc ii}] 6583$\AA$ emission line in steps of 140 
km s$^{-1}$ with a resolution of 240 km s$^{-1}$. The first four frames have the red 
side dominated by H$\alpha$ emission (the first two extremely so and whose {\sc shape} models 
have been left out for this reason) and the final frame is a sum of all those previously overlaid 
with the red-blue model distribution. Influence of the H$\alpha$ emission is visible in the 
form of the persistent S/SE feature  up to frame 7.}
\label{fig:channelmaps}
\end{figure*}

\begin{figure}
\centering
\begin{minipage}{\columnwidth}
\centering
\includegraphics[width=8cm]{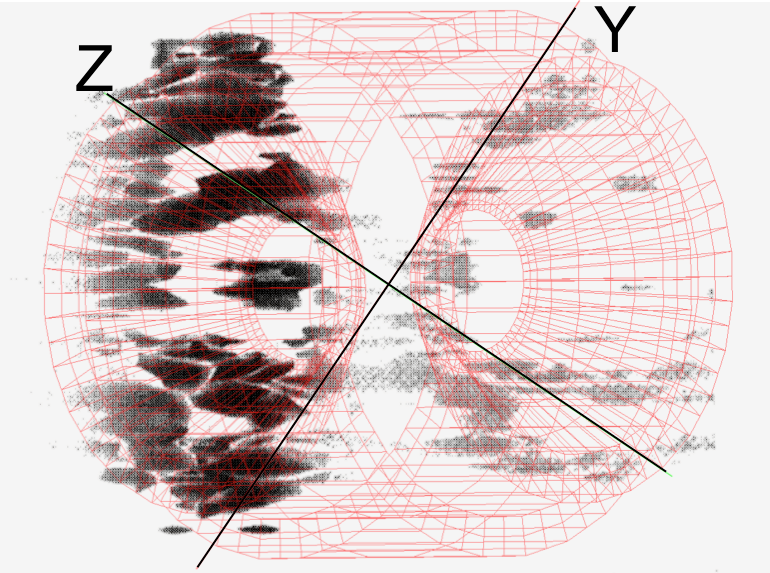}
\caption{ Z cut model presented by \cite{lawrencefp} (greyscale, adapted from their  Fig. 6, panel C) with our model overlaid on the observations (the mesh). Here we see the model polar features trace out the morphology well.}
\label{fig:channelmaps_3dfits}
\end{minipage}
\end{figure}

\begin{figure}
\centering
\begin{minipage}{\columnwidth}
\centering
\includegraphics[width=8cm]{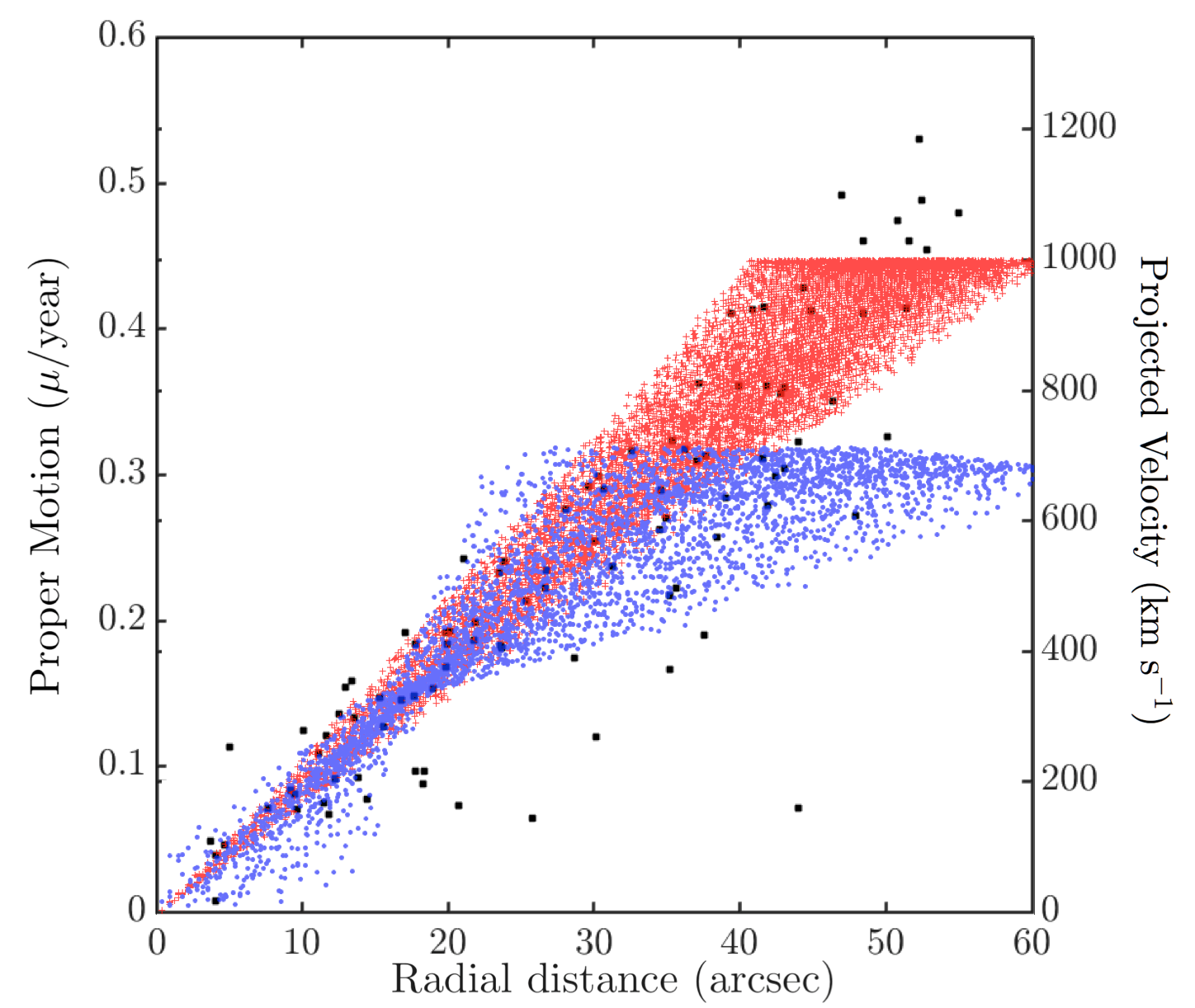}
\caption{Fit of the model presented here to the proper motion versus radial distance. The filled black squares are proper motion measurements from the online data table of L12 in terms of radial distance. The deviation from pure radial expansion is accounted for by the polar cones (blue overlay). The red corresponds to the cylinder fit to data presented in L12.}
\label{fig:PM_fit}
\end{minipage}\hfill

\begin{minipage}{\columnwidth}
\centering
\includegraphics[width=8cm]{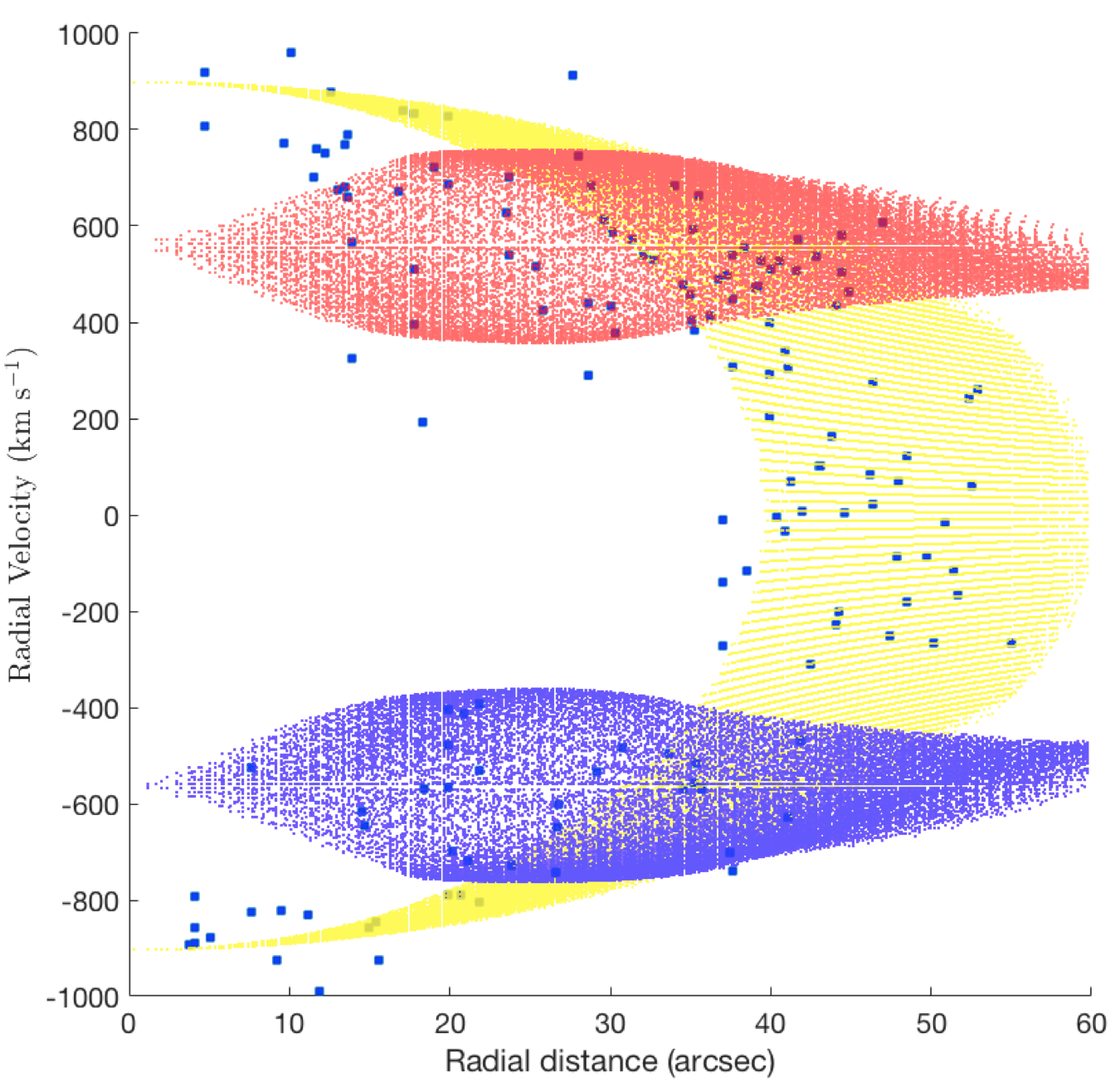}
\caption{Equivalent fit to  Fig. 9, upper panel, in L12. The yellow represents the main cylindrical shell, whereas the red and the blue cover the polar features.}
\label{fig:L12fig9}
\end{minipage}
\end{figure}

Previous efforts towards the understanding of the shell structure were unable to derive the nova shell's true 
morphology; the shell is almost consistently referred to as roughly circular and asymmetric. The flattened SW part of the shell has been given much attention 
since it is the location of interaction with pre-existing material; however, previous authors failed to address the flattened NE part of the shell, although it is remarked upon in several works, e.g. \cite{lawrencefp}. L12 argues for the 
eventual circularisation of the shell; however, here we find that the main shell most probably consists of a barrel-like equatorial structure (Fig. \ref{fig:explines}; Table \ref{cylinderfeatures}) 
and will therefore not experience such a progression towards spherical symmetry. 
The position angle derived for the nova shell here fits that of the fossil bipolar planetary nebula \citep{scottbip}. 
In the literature \cite{Seittermorph} provided a symmetric model for the morphology by using a similar methodology 
to that employed here. Their model has a prolate structure, but we find an axial ratio closer to 1:1. More importantly we find that the equatorial and polar regions are perpendicular 
to the orientation suggested by \cite{Seittermorph}, \cite{Seaquist89}, and \cite{anuprabhu}. The newer data, however, firmly indicates that this previous model cannot fit the Doppler 
overall distribution of the knots or channel maps 
of \cite{lawrencefp}, see Fig. \ref{fig:channelmaps} for the fit of our model to their channel maps and Fig. \ref{fig:channelmaps_3dfits} for an overlay of our model on their z projection of the shell. 
\cite{lawrencefp} interpreted the structure of GK Per as $\lq\lq$approximately spherical (with) striking deviations in the symmetry.'' The aforementioned deviations were (i) a northern blue-shifted region, (ii) a low-radial-velocity $\lq\lq$central'' region and (iii) a small bulge to the south-southwest. These apparent irregularities can be explained within the context of our model as (i) the blue-shifted face of the cylinder, (ii)  the northern and southern sides of the main barrel shell, and (iii)  a combination of the eastern-polar-feature (blue-shifted knots) combined with some knots attributable to the main shell (those that are red-shifted).  A comparison of the fit of two separate proper motion studies of the knots associated with the shell (L12 and S12) can be seen in   Fig. \ref{fig:PM_fit}, and a comparison of the radial velocity measurements of L12 to our model can be seen in Fig. \ref{fig:L12fig9}.

Referring to Fig. \ref{fig:explines}, several knots possibly associated with the western polar 
region of the oblate structure (i.e. along the minor, NW-SE, axis of the cylinder) have higher 
expansion velocities than our model suggests, which is a likely analogue to the diametrically opposed feature (iii) mentioned above. These features could be due to recent 
significant interaction of these knots with preexisting circumstellar material which is enhanced 
in this direction (see Fig. \ref{maywise} bottom panel). Finally, the inhomogeneous nature of 
the circumstellar matter evident in Fig. \ref{maywise} gives a natural explanation as to why some knots 
in this direction are not significantly slowed and appear as outflows with large spatial displacements in 
Fig. \ref{fig:explines}(c).  Their displacement may also be apparent as the majority of such knots have substantial 
errors in their distance determinations.

\subsubsection{Criss-cross mapping}

Criss-cross mapping is a relatively new technique presented in \cite{crisscross} for the first time, see also \cite{akrascross}. The technique leads to clues in the analysis of internal proper motions of gaseous nebulae by seeing where the proper motion vectors converge when stretched to infinity. 

\begin{figure*}
\centering
\includegraphics[width=16cm]{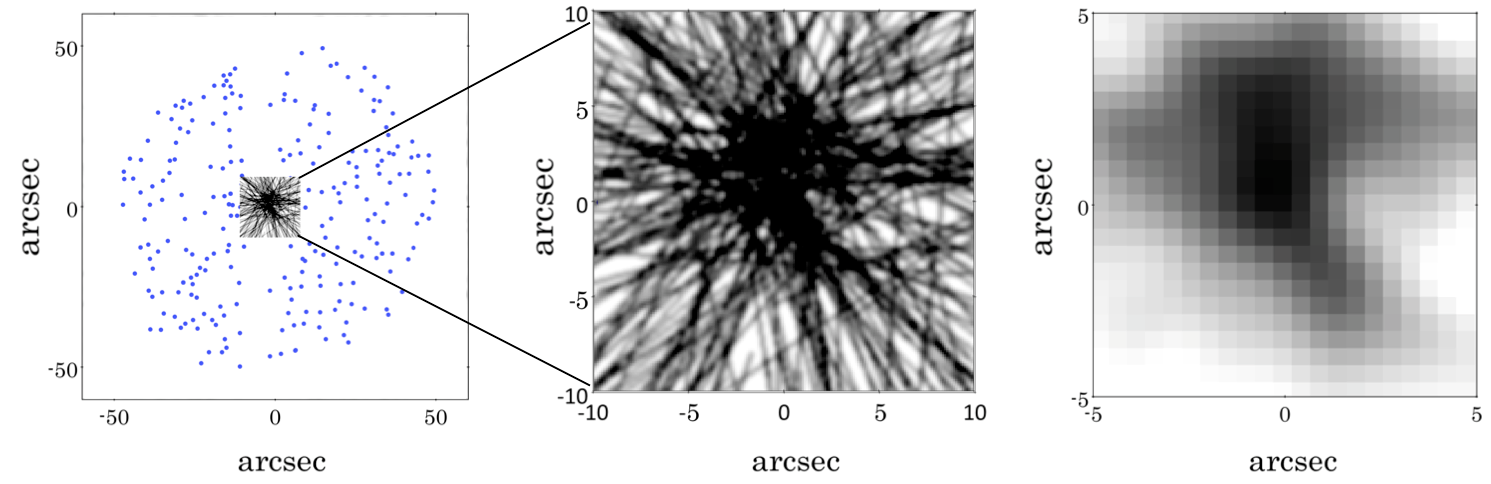}
\caption{Criss-cross mapping technique applied to GK Per observations. The criss-cross map shows a radial outflow with a discernible shift 2" to the north, although  within the noise from the observations.}
\label{fig:crisscross}
\end{figure*}

Applying this technique to the proper motion measurements in the online data table of L12, see Fig. \ref{fig:crisscross}, an apparent offset of about 2" to the north of the geometric centre of the nova shell is observed. Apart from the nice radial outflow seen in these maps, pairing Fig. \ref{fig:crisscross} with the proper motion measurement of the central star of 0.015" year$^{-1}$  \citep{Bode04},  a kinematical age of 130 years is obtained for the offset of the geometrical centre in good agreement with the age of the shell. However, these results lie within the observational noise and should be taken with caution. Another interpretation for the geometrical offset is that the nova shell is partially bipolar and that the expansion is not exactly radial at mid-latitudes.

\subsection{Kinematics and imaging of the jet-like feature}
 
Deep imaging reveals a faint jet-like feature \citep{anuprabhu} which protrudes from the NE of the nebula and stretches out into finger-like structures until it eventually disappears, see Fig. \ref{maywise}, top panel. 
Previously there have been several theories for the origin of this enigmatic feature, \citep{Anupama:aa,Bode04,Shara:2012aa}.  Light-echo contours over 
IRAS imagery \citep{Bode04} hinted that it predated the nova shell. Shara et al. (2012) discussed 
three possible origins for this feature including a Mira-like tail (which they finally rule out), a collimated jet, and a rotating feature.

\begin{figure}[h!]
\centering
\includegraphics[width=6.5cm]{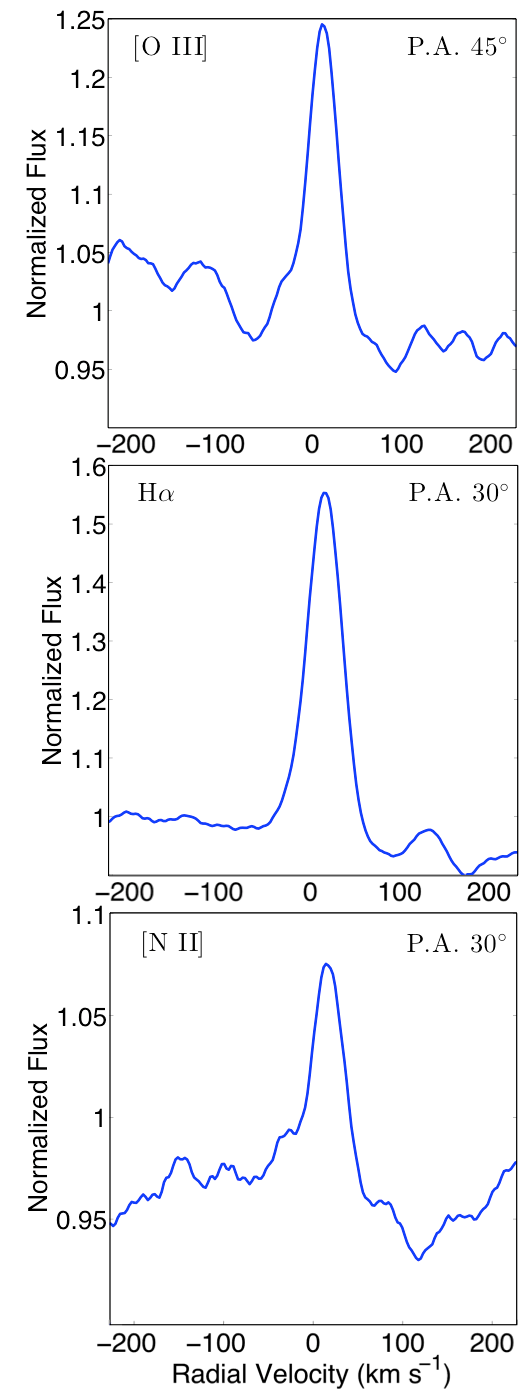}
\caption{Top panel: Resolved [O~{\sc iii}] 5007$~\AA$ line of the jet-like feature taken on 2014-Nov-28 
at a P.A. = 30$^{\circ}$, as illustrated in Fig. \ref{fig:slits}. The two lower panels are of the H$\alpha$ 6563$~\AA$ and 
[N~{\sc ii}] 6583$~\AA$ spectral lines from a slit placed at P.A. = 45$^{\circ}$. These observations are summarised in Table \ref{observations}(a).}
\label{oiii_jet}
\end{figure}

New observations presented here, corrected for the barycentric and heliocentric velocities towards the source, give averaged radial velocities over three intersecting slit position angles of 14, 14, 16, and 17 {\raisebox{.2ex}{$\scriptstyle\pm$}} 22 km s$^{-1}$ for the [O~{\sc iii}] 5007$~\AA$, [N~{\sc ii}] 6548$~\AA$, H$\alpha$ 6563$~\AA$, and [N~{\sc ii}] 6583$~\AA$ emission lines, respectively. 
The full width at half maximum for each emission line is around 30 km s$^{-1}$, see Fig. \ref{oiii_jet}.
Since our new kinematic data strongly suggest a low velocity for the jet, we put forward the idea that the feature may be an illuminated section 
of the ancient planetary nebula. A well-explained nebula could naturally account for the curvature of the jet-like feature. Low-velocity features are regularly seen at the waist regions of extreme bipolar planetary nebulae,
e.g. NGC 2346 \citep{lowvelwaist}. Evidence for the association of this feature with the planetary nebula shell is seen in the faint enhancement of the H$\alpha$ and {[O~{\sc iii}] 5007$~\AA$} lines at the interaction region to the SW of the nova shell, as noted by \cite{Anupama:aa}, \cite{balogel}, and \cite{gkchandra15}. In further support of this hypothesis \cite{Anupama:aa} find emission extending from 20 km s$^{-1}$ to -25 km s$^{-1}$ from the 21cm HI line, with emission near -5 km s$^{-1}$ west of the nova shell corresponding to the shell's $\lq$blue' side. Hinting that the $\lq$jet' belongs to the red-shifted eastern bipolar lobe of the ancient surrounding planetary nebula.

\begin{figure}
\includegraphics[width=8cm]{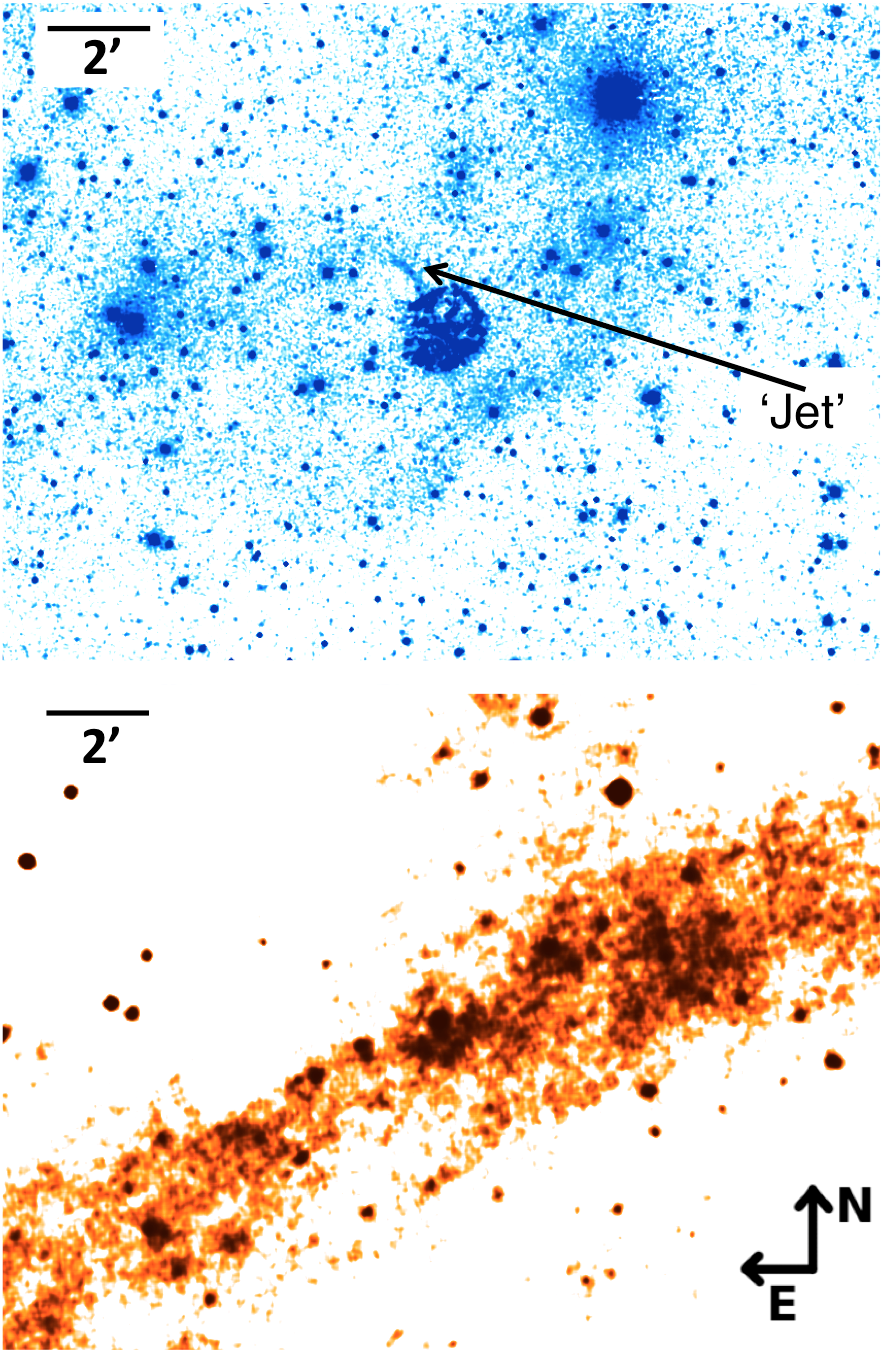}
\caption{Top panel: Binned and stretched H$\alpha$ image from the Mayall telescope where the jet-like feature is clearly visible to the NE, while the surrounding planetary nebula can also be seen. Bottom panel: WISE band 3 (resolution = 6.5$^{\prime\prime}$) illustrates the IR emitting material with which the optical shell is recombining. Strong outflow in the polar directions derived from the model presented here can also be seen.}
\label{maywise}
\end{figure}

\section{Discussion}
\label{discussion}

The newly derived morphology of the nova shell in this work is consistent with observations of the GK Per system.
One of the main findings in L12 was that there was no significant deceleration of the knots in the SW quadrant of the nova shell, 
yet it had been long believed that the shell is experiencing a stronger interaction with circumbinary material in this quadrant \citep{Seittermorph}, although L12 found a higher kinematical age for the NE part of the nova shell. 
The inclination (54$^{\circ}$) and P.A. = 120$^{\circ}$
proposed here for the nova shell pose a problem if they are related to the position angle of the jet (P.A. {\raise.17ex\hbox{$\scriptstyle\sim$}} 30$^{\circ}$) or  related to the orientation of the underlying binary.  
Even if we dismiss the P.A. for the shell in this work and consider the work of \cite{bianorbT}, where 
they derived an inclination angle of 66$^{\circ}$ for the accretion disk, then if the jet was indeed 
launched by the accretion disk  larger radial velocity measurements would be expected 
rather than those of a typical planetary nebula. Concluding that the jet-like feature must simply be an illuminated part of the ancient planetary nebula. 
A contrast in the shape of the planetary nebula and nova shell would lend valuable clues to the 
efficiency of the underlying shaping mechanism. 

As there has been some discussion on the effect of the surrounding pre-existing material on the shell morphology, the 
bullet crushing time \citep{redman_knots,bulletcrush} was calculated based on density ratios and clump properties, see Fig. 9. 
The bullet crushing time is the hydrodynamical timescale of an individual knot $\rm{t{_d}}$ and is given by $\chi$, the density ratio of the circumbinary material to that of an individual clump, the clump radius $\rm{R_c}$, and the velocity of the clump moving through the local medium is $\rm{V_s}$ : 

\begin{equation}
\rm{t{_d}} = \chi {\dfrac{\rm{R_c}}{\rm{V_s}}}. 
\end{equation}

\begin{figure}[h!]
\includegraphics[width=10.5cm]{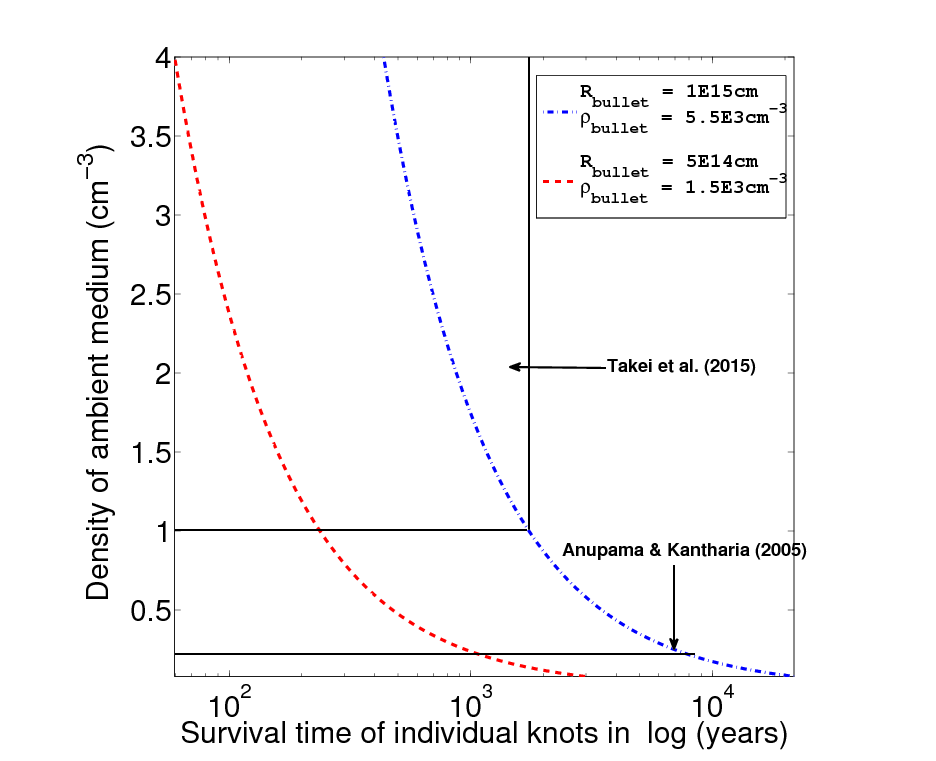}
\caption{Bullet crushing time calculations. The blue line (dot-dashed) shows the maximum length of time, depending on density, that an individual clump would remain in motion given observed constraints on the system. The red line (dashed) represents the shortest amount of time.  The area of maximum likelihood is the area between the curves.}
\label{bullet}
\end{figure}

The density estimates of the circumbinary material of 
two different works have been considered \citep{Anupama:aa,gkchandra15}. The radius of a $\lq$bullet' or knot was taken to vary between 
5x10$^{14}$ to 1x10$^{15}$ cm and the density from 1.5x10$^{3} $cm$^{-3}$ to 5.5x10$^{3} $cm$^{-3}$, following S12 and \cite{verrobsc}, see Fig. \ref{bullet}. 
Density estimates from \cite{gkchandra15} lead to hydrodynamic timescales of 333 to 1965 years, whereas those of \cite{Anupama:aa} vary from 1775 to 8239 years. Using the densities mentioned above, we can estimate the hydrodynamic
timescale according to the velocity half-lives of S12 (58 years and 220 years) and of  L12 (100 to 6000 years). L12 made their velocity half-life calculation based on a value for the initial expansion velocity of 1340 km s$^{-1}$ from \cite{pottasch} and compared it to their own observations. However, there is a range in the initial expansion velocity values for GK Per from 1240 to 1700 km s$^{-1}$ (\citealt{Anupama:aa} and \citealt{Seaquist89}, respectively) rendering this type of treatment subject to large uncertainties. Destruction of a knot follows several crushing times \citep{massflows}. Given the uncertainties, it is not possible to determine whether the main body of knots have been slowed or continue to expand more or less freely. However, our analysis suggests that some knots in the polar directions, where the densities are higher, have been slowed enough to significantly shape the nebula. Future monitoring should yield further details of the evolving knot interaction pattern, see bottom panel of Fig. \ref{maywise}.


Over time a knot will undergo several mass changing and shaping effects due to its environment such as photo-evaporation and thermal evaporation. Hydrodynamic ablation can occur via a shock transmitted through a clump; on reaching the end of the clump a strong rarefaction is reflected leading to expansion downstream that is accompanied by a lateral expansion \citep{massflows}. Lateral expansion from hydrodynamic ablation gives values of the order of 14\% of the velocity of a knot away from the GK Per central system, which is caused by the high pressure in the knot
versus the lower pressure of the surroundings and should be uniform along the length of the tail. The knot tails may also be experiencing Kelvin-Helmholtz instabilities. However the knots of GK Per are distorted on a longer wavelength and amplitude than would be expected from Kelvin-Helmholtz instability alone.

In a clump-wind interaction their relative velocities must be considered as it is believed to be the main 
mechanism in the shaping of the clumps. Under more uniform conditions, subsonic clumps have long tails and their supersonic 
counterparts display short stubby tails \citep{pittard_a}. There are a variety of tail shapes present in GK Per suggesting diverse local flow conditions. Nevertheless, we suggest that the tail shapes are due to winds external to the clumps, specifically that they are due to dwarf nova winds.
Several knots that appear to be experiencing an interaction with a following wind can be identified in the HST images. 
Five of the best examples were selected that had a sinusoidal tail shape of similar wavelength and amplitude along the outer edge of the shell such that inclination assumptions of individual knots are avoided. 
These wavy tails can be attributed to shaping by the dwarf nova winds that are thus found to 
have a velocity of {\raise.17ex\hbox{$\scriptstyle\sim$}} 4400 km s$^{-1}$. However, it should be noted that the dwarf nova outbursts 
vary in peak magnitude, meaning different energetics leading to different outburst velocities. \cite{dngkper86} gave a velocity estimate of the dwarf nova winds associated with GK Per 
of a few 1000 km s$^{-1}$. Evidence for these wavy tails can be seen in the P-V arrays in velocity space in Fig. \ref{fig:P-Vssingles}, where a sinusoidal change in velocity can be seen while looking down the length of the knot. Simply considering the time to traverse the system the dwarf 
nova responsible for the shaping of the tails in 1997 would be from {\raise.17ex\hbox{$\scriptstyle\sim$}}1969. A better understanding of the interaction between knots and the dwarf nova winds could be achieved with detailed hydrodynamical simulations equivalent to 
\cite{pittard_a,pittard_b} and\cite{mhdshock} where alternative hydrodynamical mechanisms such as the Kelvin-Helmholtz instability could be explained, or even possibly the Richtmyer-Meshkov instability since the dwarf nova episodes may be accelerating ejecta and the clump surfaces are irregular.

The southern bar which appears to be interacting with the shock has been the site of the strongest optical emission ever since the first image of the nebulosity by \cite{Barnard}. The careful examination of individual knots by L12 gave a mean weighted kinematical age of 118 $\pm$ 12 years for the nova shell (compared to the shell at 103 years old in 2004), with no hint of directional dependancies. In contrast to the work 
carried out by L12, \cite{Duerbeck} proposed that the GK Per nova shell was decelerating by 10.3 km s$^{-1}$ yr$^{-1}$ and calculated the value for the expansion of the shell at 1200 km s$^{-1}$. 
The fastest knot found using the formalisms put forward by L12 is 1190 km s$^{-1}$ and the lowest expansion velocity derived is 267 km s$^{-1}$. 
It is interesting to note that all of the knots that exceed 1000 km s$^{-1}$ 
(around 7$\%$ of all knots measured) are located on the outer limb. The knots below 600 km s$^{-1}$ (13$\%$) all reside inside the nebular boundary. 
There are only two recorded exceptions to the latter in the NE and whose P-V arrays show them to be interacting, e.g. Fig. \ref{fig:P-Vscomp} top row and Fig. \ref{fig:RV_profs} NE knot. 
The projected appearance of knots along the barrel morphology means some knots will be orientated perpendicular to our 
line of sight, as we see in the NE, e.g. Fig. \ref{fig:explines}(a), or the corresponding blue side of the shell where lack of emission to the NW can be seen.

An unexpected result here is that the derived morphology has a less extended axial ratio than expected, which goes against the grain of some of the shapes of recently modelled nova shells, e.g. RS Ophiuchi, V2672 Ophiuchi, and KT Eridani
\citep{shapeRSoph,shape_munari_oph,shape_kteri}. \cite{lloydshaping} used a 2.5D hydrodynamic code to investigate remnant shaping for a variety of 
speed classes and produced  rings, blobs, and caps as expected, but also created oblate remnants. Later, \cite{porterasphericity} included the effects of 
a rotating accreted envelope; surprisingly, the first panel in their Fig. 2 bares quite a resemblance to the morphology derived for GK Per in this work, and to model A used by \cite{Ribeiro2011}. 
Emission in Band 3 of the WISE data (bottom panel of Fig. \ref{maywise}) suggests a large amount of material in the polar regions, possibly having a significant affect on the velocity of material moving in the polar directions.

\section{Conclusions}
\label{conclusions}

In this work different axisymmetric models were considered in order to put fusrther effort towards 
the understanding of the complex morphology of the old nova shell associated with GK Per, where we find a barrel equatorial feature with polar cones to give the best fit.  In the light of our new echelle spectra gathered in November 2014, March
2015, and the WISE data archives, a new hypothesis on the origin of the mysterious jet-like feature 
has been put forward, i.e. that it is part of the ancient planetary nebula. 

We propose that the wavy tails of knots in the nova shell may be due to shaping by dwarf nova winds. 
This allows us to derive dwarf nova wind velocities of about 4400 km s$^{-1}$, although sophisticated hydrodynamical simulations would be needed to test the robustness of this hypothesis. 
Based on Doppler map profiles of the overall distribution of the nova shell knots it was found that a spherical shell, 
warped or otherwise, does not adequately explain the red-blue spread found in observations discussed here,  although a spherical shell is a good first approximation to the overall shape of the knot distribution.
Instead a cylindrical 
form with an axial ratio close to unity is found to fit best, and also shows a remarkable resemblance to the imaging data. 
After the application of the cylindrical shape to the main body of the nova shell, polar features
were included to account for the large number of knots not explained by the barrel.

The jet-like feature is most likely part of the surrounding planetary nebula owing to its low observed velocity and structure. Following the emission lines in our new observations through the shell, they 
are enhanced at a lower velocity at the area of interaction as observed in radio and X-ray observations. From the spatial modelling conducted it cannot be said whether the surrounding planetary nebula is indeed bipolar or cylindrical 
in structure with the polar over-densities (e.g. WISE band 3) attributable to both scenarios. Deep high-resolution echelle spectroscopy is needed to decipher between the two scenarios and to then test the 
efficiency of the shaping mechanisms.

%
 
\begin{acknowledgements}

The authors would like to thank the staff at the SPM and Helmos observatories for the excellent support received 
during observations. The Aristarchos telescope is operated on Helmos Observatory by the Institute for Astronomy, 
Astrophysics, Space Applications and Remote Sensing of the National Observatory of Athens. 
Based upon observations carried out at the Observatorio Astron—mico Nacional on the Sierra San Pedro M‡rtir (OAN-SPM), Baja California, MeŽxico. This research uses services or data provided by the NOAO Science Archive. NOAO is operated by the Association of Universities for Research in Astronomy (AURA), Inc., under a cooperative agreement with the National Science Foundation.
Based on observations made with the Nordic Optical Telescope, operated by the Nordic Optical Telescope Scientific Association at the Observatorio del Roque de los Muchachos, La Palma, Spain, of the Instituto de Astrofisica de Canarias.
The data presented here were obtained in part with ALFOSC, which is provided by the Instituto de Astrofisica de Andalucia (IAA) under a joint agreement with the University of Copenhagen and NOTSA.
Some of the data presented in this paper were obtained from the Mikulski Archive for Space Telescopes (MAST). STScI is operated by the Association of Universities for Research in Astronomy, Inc., under NASA contract NAS5-26555. Support for MAST for non-HST data is provided by the NASA Office of Space Science via grant NNX09AF08G and by other grants and contracts.
This publication makes use of data products from the Wide-field Infrared Survey Explorer, which is a joint project of the University of California, Los Angeles, and the Jet Propulsion Laboratory/California Institute of Technology, funded by the National Aeronautics and Space Administration.
E. Harvey wishes to acknowledge the support of the Irish Research Council for providing funding for this project under their postgraduate research scheme. S.A. gratefully acknowledges a postdoctoral fellowship from the Brazilian Agency CAPES (under their program: ``Young Talents Attraction''- Science Without Borders; A035/2013).
The authors greatly benefitted from discussions with W. Steffen, J. Meaburn, M. Lloyd, T. Jurkic, C. Neilson, and V.A.R.M. Ribeiro. We also wish to thank the anonymous referee for helpful and thoughtful comments and suggestions.

\end{acknowledgements}
%
%

\clearpage
\begin{appendix}
\section*{Appendix: Data from new observations presented in this work}
\begin{table}[]
\centering
\caption{ X and y positions of knots normalised to 2007 NOT data epoch for easier comparison with L12. The positional 
matching errors are discussed at the end of Section 2.1 and
the positional matching of knots between the two WCS matched data sets was done by matching knot shapes and centroids with consideration 
of their expected motion in the plane of the sky. Given that additional positional matching errors may arise from flux variations along a single knot between the two epochs 
(see S12, Fig. 10) 
or misidentification of a knot, average errors are of the order of $\pm$1" and maximum errors are expected to be up to 3".
The errors in the radial velocity are of the order of $\pm$22 km s$^{-1}$. As the MES wavelength range does not fully cover the three [N~{\sc ii}] 6548$~\AA$, H$\alpha$ 6563$~\AA$, and [N~{\sc ii}] 6583$~\AA$ emission lines, the following measurements were done with the [N~{\sc ii}] 6583$~\AA$ where possible, and the [N~{\sc ii}] 6548$~\AA$ elsewhere;  the H$\alpha$ 6563$~\AA$ line velocities were also measured  as a sanity check.}
\label{}
\begin{tabular}{lllllll}
x & y & Radial Velocity  & Feature  \\
arcsec & arcsec & (km s$^{-1}$) &  \\
 -32.8 & 29.0 & 137 &  barrel \\
 -31.8 & 23.8 & 450 & barrel \\
 -30.5 & 32.5 & -20 & barrel \\
-30.5 & 26.5 & -29 & barrel \\
-30.3 & 20.6 & 465 & barrel \\
-30.3 & 15.5 & 658 & barrel \\
-28.5 & 4.0 & 658 & barrel \\
-27.4 & -3.9 & 888 & barrel \\
-24.9 & -15.2 & -561 & pole \\
-24.0 & -21.3 & 574 & barrel \\
-23.2 & -24.4 & -523 & pole \\
-22.7 & -13.8 & -768 & pole \\
-21.3 & -34.5 & 485 & barrel \\
-20.8 & 35.6 & -55 & barrel \\
-19.6 & -35.9 & -440 & barrel \\
-19.1 & -33.2 & 528 & barrel \\
-18.8 & -36.5 & -653 & pole \\
-16.1 & -47.0 & -113 & barrel \\
-10.0 & -18.3 & -690 & barrel \\
-7.8 & 12.8 & -532 & barrel \\
-7.0 & -12.9 & -110 &  barrel \\
-3.7 & -6.7 & 64 & barrel \\
 -2.7 & 47.8 & -108 & barrel \\
-0.3  & 46.2 & 64 & barrel \\
 13.6 & 34.4 & -29 & barrel \\
 15.8 & 31.3 & 392 & pole \\
19.0  & 29.6 & 580 & pole \\
30.9  & 17.5 & 653 & pole \\
35.2 & 12.8 & -729 & barrel \\
39.2 & 8.0 & 498 & pole \\
40.6 & 6.0 & 290 & pole \\
44.1 & 3.5 & -222 & barrel \\
47.4 & 0.1 & -273 & barrel \\
        \end{tabular}
\end{table}
\end{appendix}
%

%


%
\bibliographystyle{aa}
\bibliography{novarefs}

\end{document}